\documentclass[a4paper,11pt,headings=big]{scrartcl}
\pdfoutput=1
\usepackage{xcolor}
\usepackage{graphicx}
\definecolor{darkblue}{rgb}{0,0,0.5}
\usepackage[colorlinks,linkcolor=darkblue,citecolor=darkblue,urlcolor=darkblue]{hyperref}
\usepackage{amsmath,amssymb}
\usepackage{bbm}
\usepackage[small]{caption}
\usepackage[normalem]{ulem}

\addtokomafont{disposition}{\rmfamily\boldmath}
\let\tmptitle\title\renewcommand{\title}[1]{\tmptitle{\LARGE #1}}
\let\tmpauthor\author\renewcommand{\author}[1]{\tmpauthor{\large #1}}
\let\tmpdate\date\renewcommand{\date}[1]{\tmpdate{\normalsize #1}}
\newcommand{\abstrct}[1]{\begin{abstract}\vspace{-2em}\small\noindent#1\end{abstract}}

\begin{document}

\titlehead{\flushright TUM-HEP-754/10}
\title{%
Viability of MSSM scenarios
at very large \texorpdfstring{$\tan\beta$}{tan(beta)}}
\author{%
Wolfgang Altmannshofer\footnote{wolfgang.altmannshofer@ph.tum.de} ~and David M.~Straub\footnote{david.straub@ph.tum.de}
\\ \normalsize\itshape
Physik-Department, Technische Universit\"at M\"unchen, 85748 Garching, Germany
}
\date{}

\maketitle

\abstrct{%
We investigate the MSSM with very large $\tan\beta > 50$, where the fermion masses are strongly affected by loop-induced couplings to the ``wrong'' Higgs, imposing perturbative Yukawa couplings and constraints from flavour physics. Performing a low-energy scan of the MSSM with flavour-blind soft terms, we find that the branching ratio of $B^+\to\tau^+\nu$ and the anomalous magnetic moment of the muon are the strongest constraints at very large $\tan\beta$ and identify the viable regions in parameter space. Furthermore we determine the scale at which the perturbativity of the Yukawa sector breaks down, depending on the low-energy MSSM parameters.
Next, we analyse the very large $\tan\beta$ regime of General Gauge Mediation (GGM) with a low mediation scale. We investigate the requirements on the parameter space and discuss the implied flavour phenomenology. We point out that the possibility of a vanishing $B\mu$ term at a mediation scale $M = 100$~TeV is challenged by the experimental data on $B^+\to\tau^+\nu$ and the anomalous magnetic moment of the muon.
}

\section{Introduction}

The Minimal Supersymmetric Standard Model (MSSM) requires two Higgs doublets $H_u$ and $H_d$ which, at tree level, couple only to up-type or down-type fermions, respectively;
\begin{equation}
W_Y^\text{MSSM} = \epsilon_{ij} \left[ Q^i Y_u U H_u^j - Q^i Y_d D H_d^j - L^i Y_\ell E H_d^j \,\right] \,.
\end{equation}

If the ratio of their two vacuum expectation values (VEVs), $v_u/v_d\equiv\tan\beta$, is large, the Yukawa couplings of the down-type fermions can be strongly enhanced with respect to their Standard Model (SM) values. This large $\tan\beta$ regime is interesting from a conceptual point of view since it explains the smallness of the bottom quark and tau lepton masses with respect to the top quark mass by a hierarchy of VEVs rather than a hierarchy of Yukawa couplings.

This possibility is well-motivated, for instance, in supersymmetric $SO(10)$ GUTs, where the top quark, bottom quark and tau lepton, sharing a common representation of the gauge group, can have a common, unified Yukawa coupling at the GUT scale $M_G\simeq 2\times10^{16}$~GeV, implying $\tan\beta\simeq50$.

Another reason to consider large values of $\tan\beta$ are the difficulties in models with gauge mediation of supersymmetry breaking to generate a $B\mu$ term in the soft SUSY breaking Lagrangian,
\begin{equation}
 \mathcal L_\text{soft} \supset -B\mu \, H_u H_d + \text{h.c.\,,}
\end{equation}
of a phenomenologically viable size \cite{Dvali:1996cu,Komargodski:2008ax}.
Indeed, if the SUSY breaking hidden sector and the visible (MSSM) sector are only connected via gauge interactions, as was suggested for a definition of General Gauge Mediation (GGM) \cite{Meade:2008wd}, the $B\mu$ term, which violates the $U(1)_\text{PQ}$ Peccei-Quinn symmetry, cannot be generated, since this $U(1)_\text{PQ}$ is preserved by gauge interactions.

While vanishing $B\mu$ would imply a vanishing VEV of $H_d$, $B\mu=0$ at the mediation scale is not preserved at low energies, since $B\mu$ is not protected by the supersymmetric non-renormalization theorem. Still, if the mediation scale is close to the electroweak scale, the radiatively generated VEV for $H_d$ is very small, i.e. $\tan\beta$ very large \cite{Dine:1996xk,Rattazzi:1996fb}. Thus the question about the phenomenologically allowed upper limit on $\tan\beta$ arises.

In the regime $\tan\beta\gg50$, the Yukawa couplings of the bottom quark and tau lepton are larger than the top Yukawa coupling, so the first concern is about the perturbativity of Yukawa couplings. Indeed, perturbativity to the GUT scale typically requires $\tan\beta\lesssim75$.
However, one should note that perturbativity all the way to the GUT scale might be a too strong requirement, since we do not know at which scale the MSSM flavour symmetry is broken, i.e. at which scale the Yukawa couplings are generated. We will therefore not insist on this requirement in this paper and simply require perturbativity up to the scale of SUSY breaking mediation.

Finally, it is well-known that, in the presence of soft SUSY breaking terms, the connection between Yukawa couplings and fermion masses is modified: loop corrections induce couplings to the ``wrong'' Higgs, which are forbidden by holomorphy in the supersymmetric limit \cite{Hall:1993gn}. Although these effects are loop-suppressed, they are very relevant for the down-type quark and charged lepton Yukawa couplings in the case of large $\tan\beta$, because they are sensitive to the large $H_u$ VEV. As a consequence, the actual values of the Yukawa couplings can be very different from the values one would expect at tree level.

Indeed, it has been pointed out recently \cite{Dobrescu:2010mk} that even for $\tan\beta\gg50$, the MSSM can be a viable theory if the dominant contributions to the down-type fermion masses arise from these threshold corrections.
In the uplifted SUSY scenario discussed in~\cite{Dobrescu:2010mk} this is realized by imposing $B\mu = 0$ at the mediation scale of SUSY breaking.
On the other hand, large values of $\tan\beta$ give rise to enhanced flavour violating processes, even with a completely flavour-blind soft sector. We therefore find it worthwhile to reconsider the phenomenological viability of the very large $\tan\beta$ region. {In particular,}
\begin{itemize}
\item {We carry out a numerical analysis of the MSSM with flavour blind soft terms at low energies for very large values of $\tan\beta$ up to 200, taking into account all the relevant and most updated constraints from the flavour sector.}
\item {We point out that in most regions of parameter space the constraint from $(g-2)_\mu$ excludes the possibility to have positive threshold corrections to the $\tau$ Yukawa coupling.}
\item {We quantify at which scale the perturbativity of the Yukawa sector breaks down for a given value of $\tan\beta$, depending on the other MSSM parameters.}
\item {We emphasize that in pure GGM, a non-zero hypercharge D-term is required to obtain very large values of $\tan\beta$.}
\item {We carry out a numerical analysis of the GGM parameter space for a fixed mediation scale of 100 TeV, showing that values of $\tan\beta$ larger than 100 are incompatible with the requirement of perturbative Yukawa couplings.}
\item {We demonstrate that for a mediation scale of 100 TeV, the uplifted SUSY scenario with $B\mu = 0$ at the mediation scale is strongly disfavoured by the flavour constraints and $(g-2)_\mu$.}
\end{itemize}

Our paper is organized as follows.
In section~\ref{sec:largetb}, we will discuss the constraints on the MSSM in the very large $\tan\beta$ region arising from the requirement of perturbative Yukawa couplings and from flavour physics.
The aim of section~\ref{sec:lowenergy} is to investigate the room for $\tan\beta\gg50$ in a general CP-conserving MSSM with flavour-blind soft terms in view of the bounds discussed in section~\ref{sec:largetb}, by means of a low-energy parameter scan.
In section~\ref{sec:ggm}, we will discuss the very large $\tan\beta$ regime in the context of General Gauge Mediation and assess whether this regime can indeed arise from the condition of a vanishing $B\mu$ term at the mediation scale. In passing, we will point out conditions on the GGM parameter space required to obtain a viable spectrum in the very large $\tan\beta$ region. {We summarize our results in section~\ref{sec:conclusions}.}

\section{The MSSM at very large \texorpdfstring{$\tan\beta$}{tan(beta)}}
\label{sec:largetb}

\subsection{Perturbativity of Yukawa couplings}

The Yukawa couplings of the third generation fermions are given at tree level by
\begin{align}
y_t &=\frac{m_t}{v \sin\beta} ~, & y_{b,\tau} &= \frac{m_{b,\tau}}{v \cos\beta} ~,
\label{eq:y3}
\end{align}
with $v\approx174$~GeV. In the large $\tan\beta$ limit, $1/\cos\beta\approx\tan\beta$ and demanding the Yukawas to be less than $\sqrt{4\pi}\approx3.54$ at a scale of, say, 1~TeV, would set an upper bound on $\tan\beta$ of roughly 250. However, as will be detailed in section~\ref{sec:tc}, the tree-level relations (\ref{eq:y3}) can be strongly modified at large $\tan\beta$, so this bound is not strict.

Irrespective of the size of low-energy threshold corrections, the bound on Yukawas at low energies is stronger if one requires them to be perturbative to some high energy scale. The RG equations for the Yukawa couplings have the general form
\begin{equation}
\frac{dy_i}{dt} = \frac{y_i}{16\pi^2 }~ \sum_{j,k}( a_i \,y_j^2 - b_k \,g_k^2 )
\end{equation}
with positive coefficients $a_i$, $b_i$, so if the Yukawa couplings are large enough (larger than their IR quasi-fixed points), their beta functions are positive, so they will grow with the renormalization scale and eventually hit a Landau pole at some scale\footnote{%
Taking into account the two-loop beta functions, the Yukawa couplings have an apparent UV fixed point instead of a Landau pole. However, this fixed point disappears once three-loop contributions are taken into account and simply signals the breakdown of perturbation theory for $y\gg\sqrt{4\pi}$ \cite{Ferreira:1996ug}.}. As a consequence, the higher the scale until where the Yukawa couplings are supposed to remain perturbative, the stronger the upper bound on their value at low energies (and consequently on $\tan\beta$).

\begin{figure}[tb]
\centering
\includegraphics[width=7.8cm]{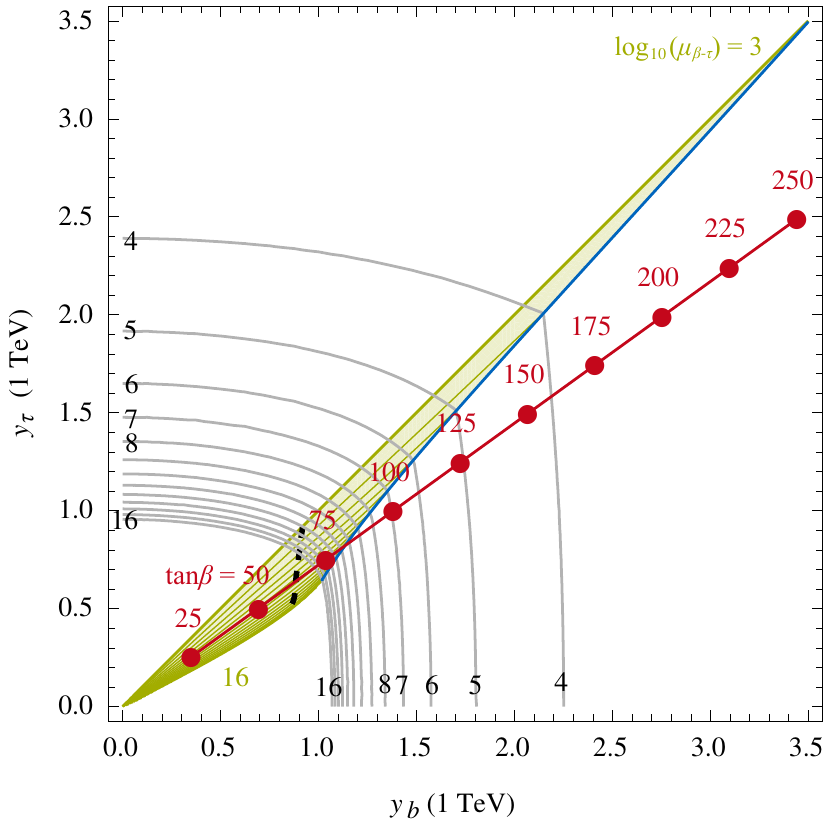}
\caption{Scale $\mu_\text{max}$ between $10^4$ and $10^{16}$~GeV at which the bottom (below the blue line) or the tau (above the blue line) Yukawa coupling exceeds the perturbativity limit $\sqrt{4\pi}$, depending on their value at 1~TeV. The gray contours show $\log_{10}(\mu_\text{max})$. The red line indicates the values of $y_{b,\tau}$ in the absence of threshold corrections, for fixed values of $\tan\beta$ between 25 and 250. The green shaded area indicates the region where perturbative $b$-$\tau$ Yukawa unification occurs between the TeV scale (uppermost green line) and $10^{16}$~GeV (lowermost green line). On the black dashed line, full $t$-$b$-$\tau$ unification takes place.
}
\label{fig:ybtaulimit}
\end{figure}

Before considering the actual low-energy values of the Yukawa couplings in the presence of threshold corrections, it is instructive to solve the RG equations for arbitrary low-energy values and determine the scale at which one of the Yukawas becomes non-perturbative. Since the first and second generation Yukawa couplings are irrelevant for this discussion and the top Yukawa is fairly insensitive to threshold corrections and independent of $\tan\beta$, one can simply   determine a scale $\mu_\text{max}$, in terms of $y_b$ and $y_\tau$ at low energies, where either $y_b$ or $y_\tau$ exceeds $\sqrt{4\pi}$. In figure~\ref{fig:ybtaulimit}, we show contours of $\log(\mu_\text{max})$ in the plane of $y_b$ and $y_\tau$, fixed for definiteness at the 1~TeV scale.

Of course, the actual low-energy values of $y_{b,\tau}$ have to be the ones reproducing the known bottom and tau masses. At the tree level, i.e. neglecting threshold corrections, they are simply given by eq.~(\ref{eq:y3}) and there is an unambiguous relation between $\tan\beta$ and the non-perturbativity scale $\mu_\text{max}$, which is shown in fig.~\ref{fig:ybtaulimit} as well. In this limit, perturbativity up to the GUT scale implies $\tan\beta\lesssim75$, for instance. However, beyond the tree level, threshold corrections will strongly modify this relation, and it will be the aim of sections~\ref{sec:lowenergy} and \ref{sec:ggm} to quantify the ``true'' $\mu_\text{max}$ for points in the MSSM parameter space.

It should also be emphasized that perturbativity up to the GUT scale is not strictly necessary, even in GUTs. To explain the peculiar hierarchies present in the Yukawa couplings, SUSY theories of flavour typically assume that the Yukawas are generated from the VEVs of dynamical flavon fields by means of the Froggatt-Nielsen mechanism \cite{Froggatt:1978nt}. While the scale at which this occurs is frequently assumed to be the GUT scale, a much lower flavour breaking scale is not forbidden. When discussing gauge mediation in section \ref{sec:ggm}, we will only require perturbativity up to the mediation scale.

A peculiar feature in the evolution of the $b$ and $\tau$ Yukawa couplings is the possibility that they unify at high energies. In $SU(5)$ and $SO(10)$ GUTs, $b$-$\tau$ Yukawa unification is a well-motivated and much-studied possibility. In figure~\ref{fig:ybtaulimit}, we also show the -- quite narrow -- region in the $y_b$-$y_\tau$ parameter space where such unification occurs at some scale between 1~TeV and $10^{16}$~GeV.
The fact that the tree-level line does not intersect the $10^{16}$~GeV contour reflects the well-known fact that $b$-$\tau$ unification at the GUT scale requires sizable threshold corrections to $y_b$ \cite{Blazek:2001sb,Blazek:2002ta,Altmannshofer:2008vr,Antusch:2008tf,Guadagnoli:2009ze,Baer:2009ie}. The Yukawa values where complete third generation Yukawa unification, i.e. $t$-$b$-$\tau$ unification, is possible, are also indicated in the figure.

\subsection{Higgs soft masses and electroweak symmetry breaking}
\label{sec:higgssoft}

Another important feature of the large $\tan\beta$ regime is its impact on radiative electroweak symmetry breaking (EWSB), by means of large Yukawa contributions to the running of the soft SUSY breaking Higgs squared masses $m_{H_{u,d}}^2$.

At the EWSB scale (which is usually assumed to be $m_\text{SUSY}=\sqrt{m_{\tilde t_1}m_{\tilde t_2}}$ to minimize logarithmic corrections to the effective potential), $m_{H_d}^2 \gtrless m_{H_u}^2$ implies $\tan\beta\gtrless1$. For $1<\tan\beta\lesssim50$, this condition can be satisfied even if $m_{H_d}^2=m_{H_u}^2$ at some high scale, since $m_{H_u}^2$ is driven towards negative values more strongly by the top Yukawa coupling compared to $m_{H_d}^2$, whose running is mainly driven by the bottom and tau Yukawas. However, for $\tan\beta\simeq50$, the third generation Yukawa couplings all become comparable, the radiative splitting mechanism becomes weaker and consequently EWSB with universal $m_{H_{u,d}}^2$ becomes more difficult \cite{Olechowski:1994gm}.

In the very large $\tan\beta$ region considered here, this problem becomes even more severe since now $y_{b,\tau}>y_t$. Therefore, $\tan\beta\gg50$ is only possible if a significant splitting $m_{H_d}^2>m_{H_u}^2$ is present already at the high scale. This significantly constrains the parameter space and immediately precludes very large $\tan\beta$ solutions in simple models like the constrained MSSM.

To understand the required magnitude of this splitting, it is useful to note that in the large $\tan\beta$ limit, one approximately has, at the EWSB scale,
\begin{equation}
m_{H_d}^2 - m_{H_u}^2 \approx m_A^2 + m_Z^2 ~,
\label{eq:DmH-1}
\end{equation}
so a heavy Higgs spectrum requires a large splitting. In addition, the value of $m_{H_u}^2$ at low energies is related to the size of the $\mu$ parameter since
\begin{equation}
- m_{H_u}^2 \approx  |\mu|^2 + \frac{1}{2}m_Z^2 ~.
\label{eq:DmH-2}
\end{equation}

\subsection{Threshold corrections to Yukawa couplings}
\label{sec:tc}

In the large $\tan\beta$ regime of the MSSM, loop induced couplings of down-type quarks and charged leptons to the up-type Higgs can lead to $O(1)$ threshold corrections to the corresponding masses~\cite{Hall:1993gn,Hempfling:1993kv,Carena:1994bv}, modify significantly CKM matrix elements~\cite{Blazek:1995nv} and also have a profound impact on flavour phenomenology~\cite{Hamzaoui:1998nu,Babu:1999hn,Carena:1999py,Carena:2000uj,Isidori:2001fv,Buras:2002vd}.
In this section we concentrate on the $\tan\beta$ enhanced threshold corrections to the bottom and tau masses. Implications for low energy flavour observables will be described in section~\ref{sec:fcnc}.

After EWSB, the tau and the bottom masses arise at tree level from the coupling to the down-type Higgs, $m_{\tau,b}^0 = y_{\tau,b} v_d$. In presence of soft SUSY breaking terms, also couplings to the up-type Higgs are generated at the loop level leading to $\tan\beta$ enhanced corrections to the tree level masses
\begin{xalignat}{2}
m_\tau &= y_\tau v_d + y_\tau^\prime v_u = m_\tau^0 (1 + \epsilon_\ell \tan\beta) ~,&\epsilon_\ell &= \epsilon_\ell^{\tilde W} + \epsilon_\ell^{\tilde B} ~,
\label{eq:tc1}
\\
m_b &= y_b v_d + y_b^\prime v_u = m_b^0 (1 + \epsilon_b \tan\beta) ~,& \epsilon_b &= \epsilon_b^{\tilde g} + \epsilon_b^{\tilde W} + \epsilon_b^{\tilde B} + \epsilon_b^{\tilde H}~.
\label{eq:tc2}
\end{xalignat}
Instead of (\ref{eq:y3}) one then has
\begin{equation} \label{eq:y3_corr}
y_\tau = \frac{m_\tau}{v} \frac{\tan\beta}{1+\epsilon_\ell \tan\beta} ~,~~~~~ y_b = \frac{m_b}{v} \frac{\tan\beta}{1+\epsilon_b \tan\beta}~,
\end{equation}
which resums the $\tan\beta$ enhanced beyond leading order corrections to all orders~\cite{Carena:1999py}.

To transparently display the main dependencies on the SUSY parameters, we now give approximate expressions for the loop factors $\epsilon_\ell$ and $\epsilon_b$ assuming all squarks and sleptons to have a common mass $\tilde m$. In addition we assume that gaugino masses $M_{1,2,3}$, the $\mu$ parameter as well as the trilinear couplings $A_{t,b,\tau}$ are real. Under these assumptions, the loop factors are well approximated by
\begin{eqnarray}
\epsilon_b^{\tilde g} = \frac{\alpha_s}{4 \pi} \frac{8}{3} \frac{\mu M_3}{\tilde m^2} f_1(x_3) ~,&&~~
\epsilon_b^{\tilde H} = \frac{\alpha_2}{4 \pi} \frac{m_t^2}{2 M_W^2} \frac{\mu A_t}{\tilde m^2} f_1(x_\mu)~, \nonumber
\end{eqnarray}
\begin{equation}
\epsilon_b^{\tilde W} = \epsilon_\ell^{\tilde W} = -\frac{\alpha_2}{4 \pi} \frac{3}{2} \frac{\mu M_2}{\tilde m^2} f_2(x_2,x_\mu) ~,
\end{equation}
with the mass ratios $x_{2,3} = M_{2,3}^2/\tilde m^2$ and $x_\mu = \mu^2/\tilde m^2$. For the loop functions one has $f_1(1) = f_2(1,1) = 1/2$ and their explicit expressions are reported in appendix~\ref{sec:loop}. Our conventions are such that $M_3$ is always real and positive, the left-right mixing entry in the stop mass matrix is given by $m_t (A_t - \mu \cot\beta)$, and the ones in the sbottom and stau mass matrices by $m_{b,\tau} (A_{b,\tau} - \mu\tan\beta)$. As the Bino contributions $\epsilon^{\tilde B}_{b,\tau}$ are parametrically suppressed by the small gauge coupling $\alpha_1$, we refrain from giving explicit expressions also for them. They can be found e.g. in~\cite{Dobrescu:2010mk}.

The dominant correction to the bottom Yukawa is typically given by the gluino contribution $\epsilon_b^{\tilde g}$ which is positive (negative) for positive (negative) $\mu$. As $\epsilon_b^{\tilde g}$ does not decouple with $\mu$, the largest corrections to the bottom Yukawa are expected when $|\mu|$ is large.
For large values of $A_t$ (and light stops), also the Higgsino contribution $\epsilon_b^{\tilde H}$ can in principle become important and, depending on the sign of $A_t$, interfere constructively or destructively with the gluino contribution.

Concerning the $\tau$ Yukawa, due to the absence of gluino and Higgsino corrections, the Wino contribution $\epsilon_\ell^{\tilde W}$ is typically dominant and its sign is determined by $-\text{sign}(\mu M_2)$.

\subsection{Low energy constraints}
\label{sec:fcnc}

The most important low energy constraints on the MSSM with large $\tan\beta$ and Minimal Flavour Violation arise from helicity suppressed observables, where SUSY contributions can be enhanced by powers of $\tan\beta$. If there are no new sources of CP violation apart from the phase of the CKM matrix then the crucial observables are the branching ratios of the $B$ decays $B\to\tau\nu$, $B_s\to\mu^+\mu^-$ and $B\to X_s\gamma$ in the flavour sector, as well as the anomalous magnetic moment of the muon as a flavour conserving observable~\cite{Isidori:2006pk}. We will now briefly review the experimental status of these observables and discuss the corresponding SUSY contributions.\footnote{While we only give simple approximate expressions for the dominant SUSY contributions, in our numerical analysis we use the full set of leading order contributions and include the $\tan\beta$ enhanced beyond leading order corrections following the general method described in~\cite{Buras:2002vd}. As the iteration procedure of~\cite{Buras:2002vd} breaks down for $\epsilon \tan\beta > 1$, we start the iteration using the initial Yukawa couplings~(\ref{eq:y3_corr}) that constitute already the correct resummed result in the decoupling limit, $v^2/m_\text{SUSY}^2 \to 0$. The iteration then serves to include corrections beyond the decoupling limit. An alternative, fully analytic method has been proposed in~\cite{Hofer:2009xb}.}

\subsubsection{The anomalous magnetic moment of the muon}

The current SM prediction for the anomalous magnetic moment of the Muon, $a_\mu = \frac{1}{2}(g-2)_\mu$, reads~\cite{Prades:2009qp}
\begin{equation} \label{eq:amu_SM}
a_\mu^{\rm SM} = (11\,659\,183.4 \pm 4.9) \times 10^{-10}~.
\end{equation}
Combined with the final experimental result from the Muon (g-2) Collaboration~\cite{Bennett:2006fi,Roberts:2010cj}
\begin{equation} \label{eq:amu_exp}
a_\mu^{\rm exp} = (11\,659\,208.9 \pm 6.3) \times 10^{-10}~,
\end{equation}
one finds a discrepancy at the level of $3.2\sigma$~\cite{Prades:2009qp}
\begin{equation} \label{eq:Damu}
\Delta a_\mu = a_\mu^{\rm exp} - a_\mu^{\rm SM} = (25.5 \pm 8.0) \times 10^{-10}~.
\end{equation}
Such a discrepancy can easily be explained in the MSSM with large $\tan\beta$ where the following approximate expression holds for the typically dominant Wino contribution~\cite{Moroi:1995yh,Marchetti:2008hw}
\begin{eqnarray} \label{eq:Damu_SUSY}
\Delta a_\mu^{\rm SUSY} &=& \frac{\alpha_2}{4\pi} m_\mu^2 \; \frac{t_\beta}{1+\epsilon_\ell t_\beta}\; \frac{\mu M_2}{\tilde m^4} f_4(x_2,x_\mu) ~,
\end{eqnarray}
with $f_4(1,1) = 5/12$ and the explicit expression for the loop function is given in appendix~\ref{sec:loop}. There are corners in the MSSM parameter space where also Bino contributions can become non-negligible. Explicit expressions for them can be found in~\cite{Moroi:1995yh}.

The sign of the Wino contribution in (\ref{eq:Damu_SUSY}) is determined by ${\rm sign}(\mu M_2)$. The current data on $(g-2)_\mu$ thus usually excludes a relative minus sign between $\mu$ and $M_2$ at more than $3\sigma$. We remind that also the sign of the dominant contribution to $\epsilon_\ell$ is determined by exactly the same parameter combination (cf. section \ref{sec:tc}), with $\epsilon_\ell$ {\it negative} for ${\rm sign}(\mu M_2) = +1$.

\subsubsection{The decay \texorpdfstring{$B^+ \to \tau^+ \nu$}{B->tau nu}}
\label{sec:Btaunu}

The tree level decay $B^+ \to \tau^+ \nu$ is a sensitive probe of models with extended Higgs sectors~\cite{Hou:1992sy}. Indeed, in the MSSM with large $\tan\beta$, its branching ratio can differ significantly from the SM prediction~\cite{Akeroyd:2003zr,Isidori:2006pk}. Tree level charged Higgs contributions interfere destructively with the SM ones and lead to
\begin{equation} \label{eq:RBtaunu}
R_{B\tau\nu}=\frac{{\rm BR}(B^+ \to \tau^+ \nu)}{{\rm BR}(B^+ \to \tau^+ \nu)_{\rm SM}} \simeq \left( 1 - \frac{m^2_{B^+}}{M^2_{H^+}} \frac{t^2_\beta}{(1 + \epsilon_0 t_\beta)(1 + \epsilon_\ell t_\beta)} \right)^2 ~,
\end{equation}
with $\epsilon_0$ defined as $\epsilon_0 = \epsilon_b^{\tilde g} + \epsilon_b^{\tilde W} + \epsilon_b^{\tilde B}$.

While the $B^+ \to \tau^+ \nu$ decay is not yet ``observed'' with a $5\sigma$ significance, the current experimental world average for the branching ratio reads~\cite{Tisserand:2009ja}
\begin{equation}
{\rm BR}(B^+ \to \tau^+ \nu)_{\rm exp} = (1.73 \pm 0.35) \times 10^{-4}~.
\end{equation}
SM predictions for the branching ratio that rely on fits of the Unitarity triangle result in values roughly $2.5 \sigma$ below the experimental data~\cite{Tisserand:2009ja,Bona:2009cj,Altmannshofer:2009ne}, leading to severe constraints on the destructively interfering charged Higgs contributions. To be conservative, we use instead as a SM prediction~\cite{Altmannshofer:2009ne} (see also~\cite{Rosner:2010ak})
\begin{equation}
{\rm BR}(B^+ \to \tau^+ \nu)_{\rm SM} = (1.10 \pm 0.29) \times 10^{-4}~,
\end{equation}
which implies
\begin{equation} \label{eq:RBtaunu_exp}
R_{B\tau\nu} = 1.57 \pm 0.53 ~.
\end{equation}

We remark that $B \to \tau \nu$ alone cannot exclude a scenario where the SM contribution is overcompensated by a charged Higgs contribution more than twice as large as itself. However, the experimental data on the $K \to \mu \nu$ and $B \to D \tau \nu$ decays exclude such a fine tuned situation~\cite{Antonelli:2008jg,Antonelli:2009ws,Bona:2009cj}.
Therefore, despite the still rather large experimental and theory uncertainties, one expects stringent lower bounds on the charged Higgs mass in the very large $\tan\beta$ regime.

\begin{figure}[tb]
\centering
\includegraphics[width=7.8cm]{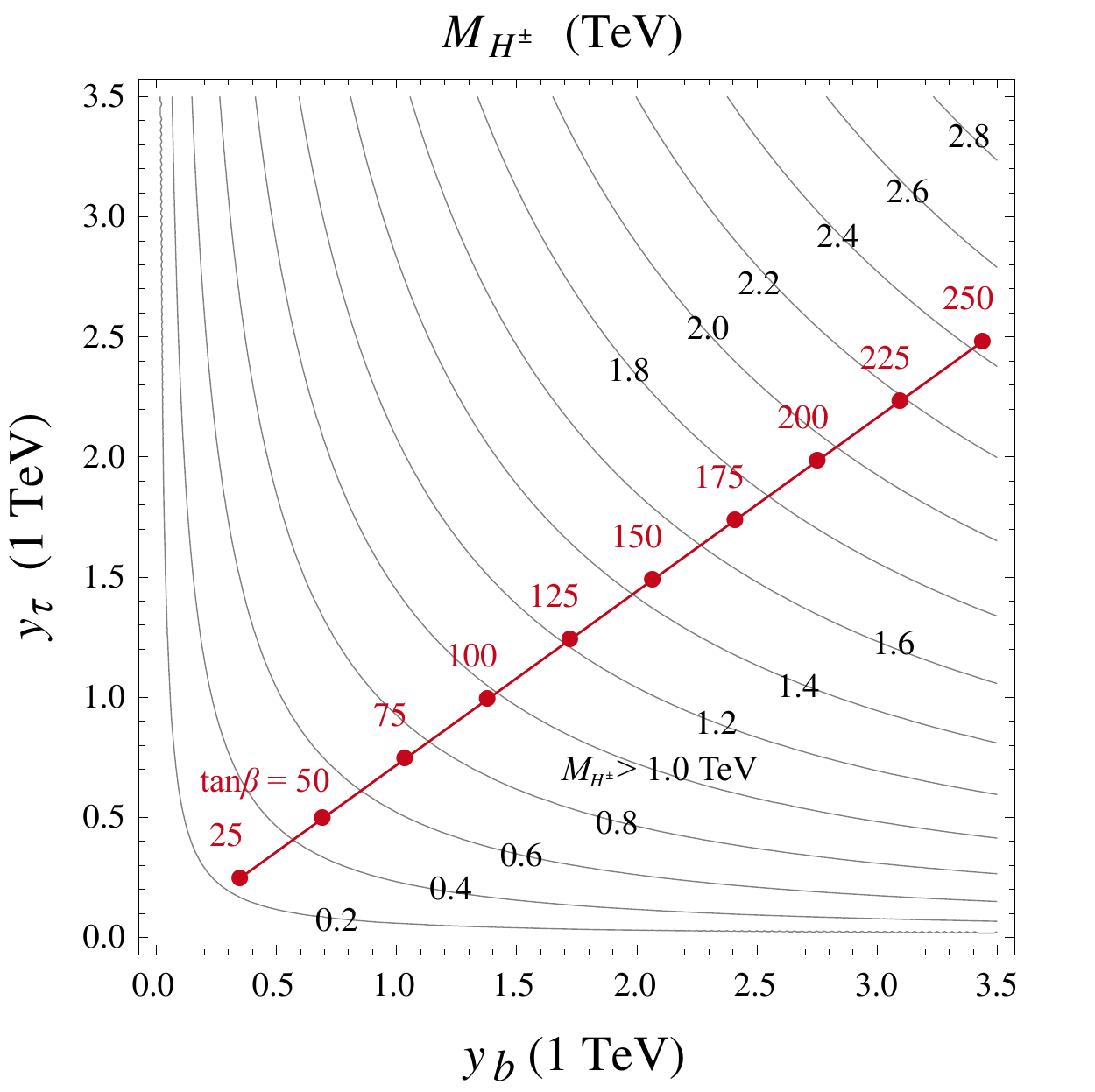}
\caption{Lower bound on the charged Higgs mass $m_{H^\pm}$ in TeV arising from BR$(B\to\tau\nu)$ at the $2\sigma$ level, depending on the values of the bottom and tau Yukawa couplings at the TeV scale.
These bounds are valid under the assumption that gluino contributions to the $y_b$ threshold correction dominate over Higgsino contributions.
The points on the red line denote the values of the Yukawa couplings at a given value of $\tan\beta$ in the absence of threshold corrections, as in fig.~\ref{fig:ybtaulimit}.
}
\label{fig:Btaunu}
\end{figure}

To illustrate this point, we show in fig.~\ref{fig:Btaunu} the lower bounds on the charged Higgs mass in the $y_b$-$y_\tau$ plane, based on applying~(\ref{eq:RBtaunu_exp}) at the $2\sigma$ level. The plot is obtained under the assumption that $\epsilon_b^{\tilde H} \ll \epsilon_b^{\tilde g}$ such that $\epsilon_b \simeq \epsilon_0$ and $R_{B\tau\nu}$ in~(\ref{eq:RBtaunu}) can directly be expressed in terms of the Yukawa couplings~(\ref{eq:y3_corr}), which will indeed be the case in the scenarios that we will consider in sections~\ref{sec:lowenergy} and~\ref{sec:ggm}.

The red line shows the Yukawa couplings in the absence of threshold corrections for fixed values of $\tan\beta$. Along this line we find
\begin{equation} \label{eq:Higgs_bound}
m_{H^\pm} > 9.6\,\text{GeV} \times \tan\beta ~,
\end{equation}
which is less stringent than the bound found in~\cite{Bona:2009cj}, due to our more conservative SM prediction. In the presence of threshold corrections, this bound can be relaxed, provided $\epsilon_b > 0$ and/or $\epsilon_\tau > 0$.

\subsubsection{The decay \texorpdfstring{$B_s \to \mu^+ \mu^-$}{B(s)->mu+mu-}}

The experimental bound on the branching ratio of the rare decay $B_s \to \mu^+ \mu^-$~\cite{:2007kv},
\begin{equation}
{\rm BR}(B_s \to \mu^+ \mu^-)_{\rm exp} < 5.8 \times 10^{-8}
\text{~~at 95\% C.L.,}
\end{equation}
is roughly a factor 15 above the SM prediction~\cite{Altmannshofer:2009ne}
\begin{equation}
{\rm BR}(B_s \to \mu^+ \mu^-)_{\rm SM} = (3.60 \pm 0.37) \times 10^{-9}~.
\end{equation}
Still, $B_s \to \mu^+ \mu^-$ constitutes a very important constraint of the MSSM at large $\tan\beta$ as the branching ratio grows with $\tan^6\beta$~\cite{Choudhury:1998ze,Babu:1999hn}. Approximately one has
\begin{equation}
R_{B_s\mu\mu}=\frac{{\rm BR}(B_s \to \mu^+ \mu^-)}{{\rm BR}(B_s \to \mu^+ \mu^-)_{\rm SM}}= \left|A\right|^2 + \left|1- A\right|^2 ~,
\end{equation}
with the dominant Higgsino contribution to $A$ given by
\begin{equation}
A^{\tilde H} = \frac{m_{B_s}^2}{8 m_A^2} \frac{t^3_\beta}{(1+\epsilon_b t_\beta)(1+\epsilon_0 t_\beta)(1+\epsilon_\ell t_\beta)} ~ \frac{m_t^2}{M_W^2} \frac{\mu A_t}{\tilde m^2} \frac{f_1(x_\mu)}{Y_0(x_t)}~.
\end{equation}
The loop function $f_1$ appeared already in the discussion of threshold corrections in section \ref{sec:tc} and for the SM loop function one has $Y_0(x_t) \simeq 0.96$.

We note that $A^{\tilde H}$ is proportional to the stop trilinear coupling $A_t$ giving the possibility to relax the constraint from $B_s \to \mu^+\mu^-$ by lowering $A_t$.

\subsubsection{The decay \texorpdfstring{$B \to X_s \gamma$}{B->X(s)gamma}}

The good agreement between the experimental data on the branching ratio of $B \to X_s \gamma$~\cite{Barberio:2008fa}
\begin{equation}
{\rm BR}(B \to X_s \gamma)_{\rm exp} = (3.52 \pm 0.25) \times 10^{-4}~,
\end{equation}
and the corresponding NNLO SM prediction~\cite{Misiak:2006zs}
\begin{equation}
{\rm BR}(B \to X_s \gamma)_{\rm SM} = (3.15 \pm 0.23) \times 10^{-4}~,
\end{equation}
leads to severe constraints on the MSSM parameter space, in particular for large values of $\tan\beta$. The prediction for the branching ratio in the MFV MSSM with no new sources of CP violation can be well approximated by the following equation~\cite{Lunghi:2006hc}
\begin{eqnarray}
R_{bs\gamma} = \frac{{\rm BR}(B \to X_s \gamma)}{{\rm BR}(B \to X_s \gamma)_{\rm SM}} &=& 1 - 2.41 \,C_7^{\rm NP} -  0.75\,C_8^{\rm NP} \nonumber \\
&&+ 1.59\,\left(C_7^{\rm NP}\right)^2 + 0.27\,\left(C_8^{\rm NP}\right)^2 + 0.82\,C_7^{\rm NP} C_8^{\rm NP}~,
\end{eqnarray}
with the NP contributions to the Wilson coefficients $C_7$ and $C_8$ evaluated at a scale $\mu_0 = 160$~GeV. The leading Higgsino contributions are $\tan\beta$ enhanced and read at the matching scale
\begin{equation} \label{eq:C78Higgsino}
C_{7,8}^{\tilde H} = - \frac{t_\beta}{1 + \epsilon_b t_\beta} \; \frac{m_t^2}{\tilde m^2} \; \frac{A_t \mu}{\tilde m^2} f_{7,8}(x_\mu)~.
\end{equation}
For the loop functions one has $f_7(1) = 5/72$, $f_8(1) = 1/24$ and their explicit expressions can be found in appendix~\ref{sec:loop}.

Even for a completely flavour blind soft sector, there are two-loop contributions involving gluinos that have been explicitly worked out in~\cite{Hofer:2009xb}. They can lead to non-negligible effects in particular in $C_8$ and read
\begin{equation} \label{eq:C78gluino}
C_{7,8}^{\tilde g} = - \frac{\alpha_s}{4\pi} \frac{t^2_\beta}{(1 + \epsilon_0 t_\beta)(1 + \epsilon_b t_\beta)} \; \frac{m_t^2}{\tilde m^2} \; \frac{\mu A_t}{\tilde m^2} \; \frac{\mu M_3}{\tilde m^2} f_1(x_\mu) \tilde f_{7,8}(x_3) ~.
\end{equation}
The loop functions are given in appendix~\ref{sec:loop}, and one finds $\tilde f_7(1) = 1/27$ and $\tilde f_8(1) = 5/36$. The suppression of these contributions by $\alpha_s/4\pi \simeq 0.01$ can be compensated by the additional factor $\tan\beta$. They are consistently included in our numerical analysis using the general procedure described in~\cite{Buras:2002vd}.

As a final remark we note that both contributions, (\ref{eq:C78Higgsino}) and (\ref{eq:C78gluino}), are proportional to the stop trilinear coupling $A_t$. Thus the constraint from $B \to X_s \gamma$ can be relaxed for small values of $A_t$.

\section{Numerical analysis: low-energy approach}
\label{sec:lowenergy}

In the last section, we outlined the features of the MSSM at very large $\tan\beta$ and discussed the threshold corrections to Yukawa couplings and the impact on FCNCs. Now, we want to quantify the room left for $\tan\beta\gg50$ by means of a low-energy parameter scan.

In view of the potentially dangerous effects on flavour and CP violating observables at very large $\tan\beta$, we restrict ourselves to a flavour-blind and CP-conserving subspace of the full MSSM parameter space. In particular, we assume
\begin{align}
\mathbf m_{Q,U,D,L,E}^2 &= m_{Q,U,D,L,E}^2  \times \mathbf 1 ~, 
\label{eq:LEpar1}\\
\mathbf A_{u,d,l} &= A_{u,d,l} Y_{u,d,l} ~,
\label{eq:LEpar2}
\end{align}
with real $m_i^2$ and $A_I$ to hold at low energies, as well as a real $\mu$ term and real gaugino masses. We do not impose the GUT relations on the gaugino masses and allow both signs for $M_1$ and $M_2$ (note that $M_3$ can always be made real and positive by a phase redefinition).

Complete flavour blindness at low energies is of course a quite restricted subspace of the full MSSM parameter space and is not an RG invariant assumption. Therefore models with non-minimal flavour violation or with extreme non-universalities, like the radiatively induced inverted scalar hierarchies necessary in models with Yukawa unification \cite{Blazek:2001sb,Blazek:2002ta,Altmannshofer:2008vr,Guadagnoli:2009ze}, are not covered by this approach.
On the other hand, it is a good approximation to models with flavour-blind mediation of SUSY breaking and a low mediation scale like the gauge mediation models we consider in section~\ref{sec:ggm}.

In addition to the parameters in eqs.~(\ref{eq:LEpar1})--(\ref{eq:LEpar2}) and the gaugino masses, also the pseudoscalar Higgs mass $m_A$, $\mu$ and $\tan\beta$ are free parameters in this setup. We scan them in the ranges
\begin{align} \label{eq:LE_ranges}
m_{Q,U,D,L,E} &\in [0, 2] ~\text{TeV} , &
A_{u,d,l} &\in [-0.5, 0.5] ~\text{TeV} , &\\
\mu &\in [-2, 2] ~\text{TeV} , &
m_A &\in [0, 2] ~\text{TeV} , &\\
M_1 &\in [-1, 1] ~\text{TeV} , &
M_2 &\in [-2, 2] ~\text{TeV} , &
M_3 &\in [0, 6] ~\text{TeV} ,
\end{align}
for fixed, large values of $\tan\beta=(50,100,150,200)$. We restrict our analysis to a limited range of the trilinear couplings to soften possible constraints coming from $B_s \to \mu^+\mu^-$ and $B \to X_s \gamma$. In fact, small values of the trilinear couplings naturally arise in the gauge mediation scenario we consider in section~\ref{sec:ggm}.

We discard points violating existing lower bounds on sparticle and Higgs masses, as well as points violating the perturbativity condition $y_{b,\tau}<\sqrt{4\pi}$ at the EWSB scale $m_\text{SUSY}=\sqrt{m_{\tilde t_1}m_{\tilde t_2}}$.

\begin{figure}[tbp]
\centering
\includegraphics[width=14cm]{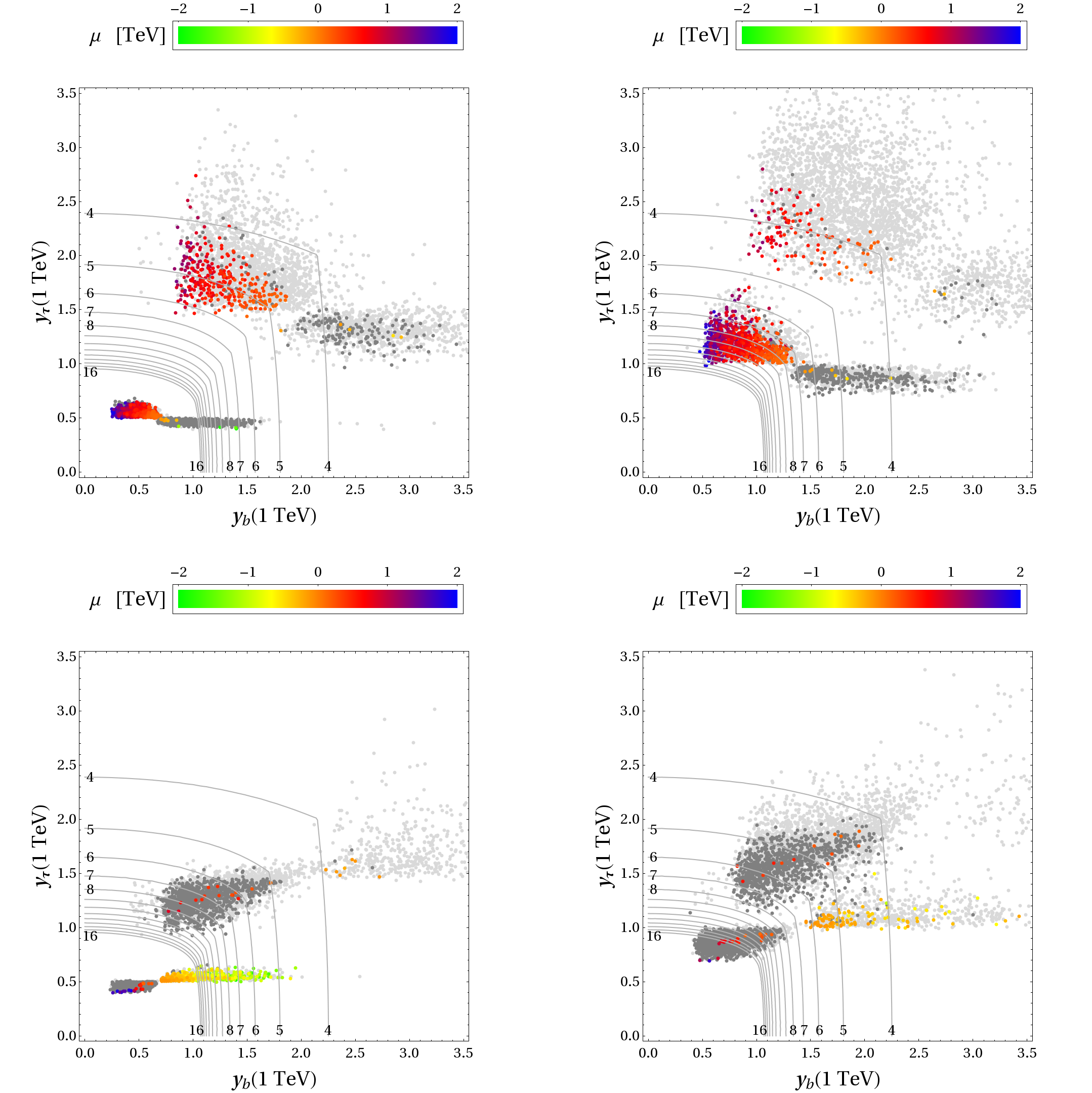}
\caption{Values of the bottom and tau Yukawa couplings at the TeV scale in the low-energy MSSM scan with flavour blind soft terms with $M_2>0$ (top row) and $M_2<0$ (bottom row). The plots in the left column show the points for $\tan\beta=50$ below the ones for $\tan\beta=150$; the plots in the right column show the ones for $\tan\beta=100$ below the ones for $\tan\beta=200$.
Light gray points are excluded at the $2\sigma$ level by the flavour constraints, while dark gray ones are excluded by $(g-2)_\mu$ alone at the $3\sigma$ level. Among the coloured points, the ones with smaller values of $\mu$ are stacked upon the ones with larger $\mu$.
The $\mu_\text{max}$ contours are as in fig.~\ref{fig:ybtaulimit}.
}
\label{fig:ybtau-LE}
\end{figure}

\begin{figure}[tbp]
\centering
\includegraphics[width=14cm]{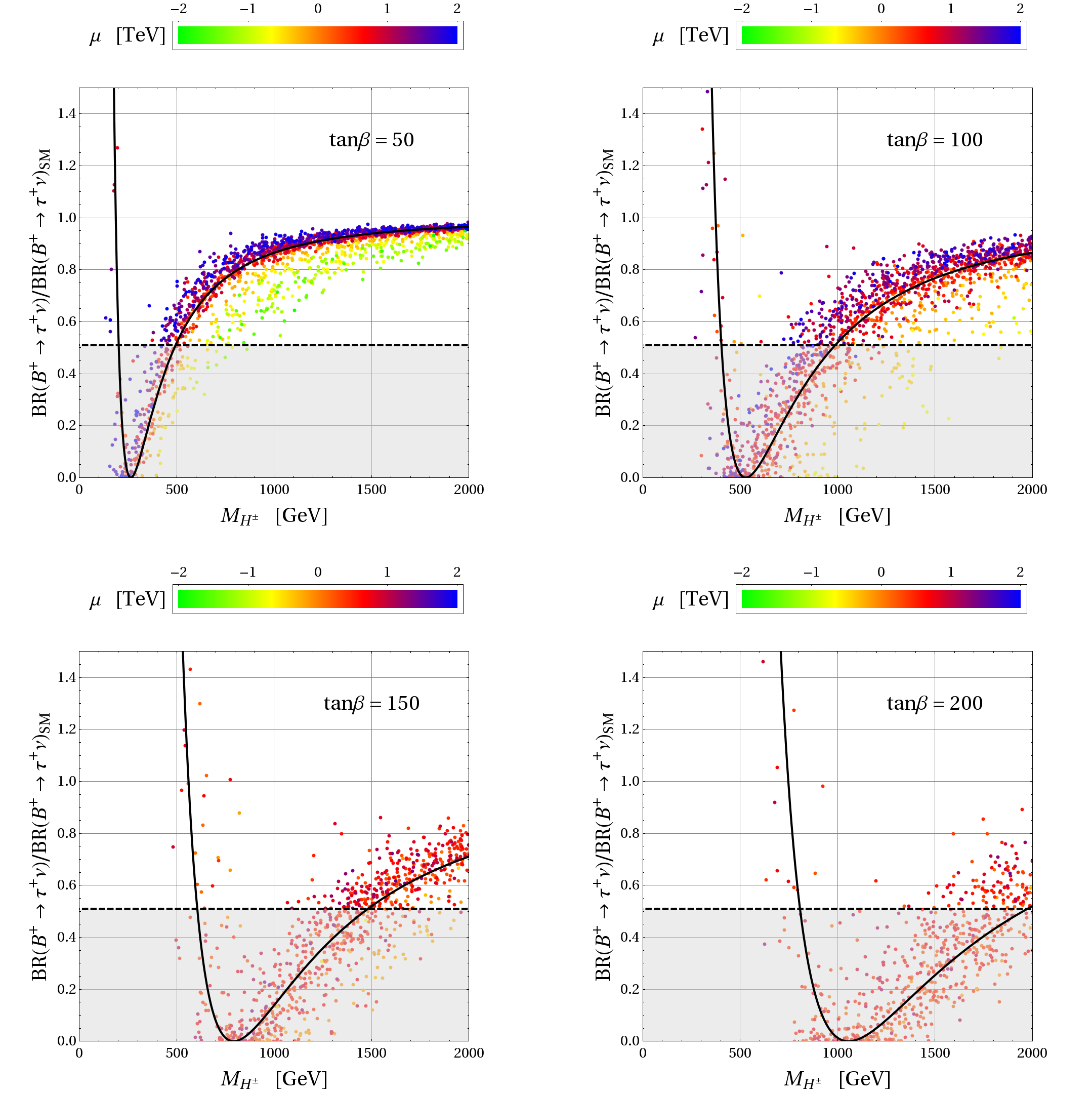}
\caption{Branching ratio of $B^+\to\tau^+\nu$ compared to its SM value as a function of the charged Higgs mass for the low-energy MSSM scan with flavour blind soft terms. All the shown points fulfill the constraints from the other flavour observables as well as from $(g-2)_\mu$. The solid black curve corresponds to the result in absence of threshold corrections and the experimental lower bound is shown as dashed line.
}
\label{fig:Btaunu_MSSM}
\end{figure}

In figure~\ref{fig:ybtau-LE}, we show the results for the bottom and tau Yukawa couplings in the presence of threshold corrections for the four different values of $\tan\beta$, superimposed on the $\mu_\text{max}$ contours of fig.~\ref{fig:ybtaulimit}.
The plots on the left (right) contain points with $\tan\beta = 50 , 150$ ($\tan\beta = 100 , 200$) and in the first (second) row the Wino mass $M_2$ is positive (negative).
Light gray points are excluded at the $2\sigma$ level by the flavour constraints, while dark gray ones are excluded by $(g-2)_\mu$ alone at the $3\sigma$ level.
The coloured points finally fulfill all imposed constraints and indicate the corresponding values of $\mu$ through the colour. We make the following observations:
\begin{itemize}
\item The four ``lobes'' of points correspond to the different signs of $(\mu,M_2)$ and meet at the $(y_b,y_\tau)$ values in absence of threshold corrections (cf. fig.~\ref{fig:ybtaulimit}). Points left of it correspond to positive threshold corrections to $y_b$ ($y_b'>0$ in the notation of eq. (\ref{eq:tc2})), points below it to positive threshold corrections to $y_\tau$ ($y_\tau'>0$), etc.
The sign of the corrections to $y_\tau$ is given by the sign of $(-\mu M_2)$, and the sign of the corrections to $y_b$ by the sign of $\mu$, as expected from the dominance of Wino contributions and gluino contributions, respectively, discussed in section \ref{sec:tc}. 
\item Almost all the points where $\text{sign}(\mu M_2)=-1$ are ruled out at more than $3\sigma$ by the $(g-2)_\mu$ constraint. This rules out positive threshold corrections to $y_\tau$, as would be preferable from the point of view of perturbativity to high scales. The best points in this respect allowed by the $(g-2)_\mu$ constraint are then the ones with positive contributions to $y_b$ and negative contributions to $y_\tau$, i.e. with $\text{sign}(\mu) = \text{sign}(M_2) = +1$.
\item There exist sporadic points with positive corrections to the tau Yukawa coupling that are in agreement with all available constraints, in particular also with the $(g-2)_\mu$. We find that these points are characterized by very large left-handed (LH) and very small right-handed (RH) slepton masses. Very heavy LH sleptons lead to a fast decoupling of the Wino contributions to $\Delta a_\mu$ and Bino contributions become non-negligible. In fact a diagram involving a Bino, a Higgsino and a RH slepton gives the dominant effect in $\Delta a_\mu$ for these points and its sign is given by $-\text{sign}(\mu M_1)$.
As the decoupling with the slepton mass is much weaker in $\epsilon_\ell$ as compared to $\Delta a_\mu$, there exist in fact corners in the parameter space where one has a dominant and positive Wino correction to $\epsilon_\ell$ and simultaneously a dominant and positive Bino contribution to $\Delta a_\mu$, provided $\text{sign}(\mu M_2) = \text{sign}(\mu M_1) = -1$.
\item The size of the threshold corrections to $y_b$ is mainly limited by the maximal possible values of $|\mu|$. Especially for very large $\tan\beta = 150,200$, $|\mu|$ typically does not exceed $1\,$TeV. The reason for that is twofold: First, very large values of $\mu \tan\beta$ lead to large left-right mixing entries in the sbottom and stau mass matrices, easily resulting in tachyonic sbottoms or staus. Second, there are sizable sbottom contributions to the lightest Higgs boson mass for large $\mu \tan\beta$~\cite{Brignole:2002bz}. These contributions are always negative and therefore decrease the lightest Higgs boson mass to values below the LEP bound for too large $\mu \tan\beta$.
\item As a side remark we note that for $\tan\beta =200$, we also find few valid points with positive corrections to the bottom Yukawa so large that $y_b$ turns out to be negative. These very fine tuned points that are not shown in fig.~\ref{fig:ybtau-LE} are characterized by large negative $\mu$ and negative $M_2$ and are clustered around $(y_b,y_\tau) \simeq (-2,2.5)$. Interestingly, the negative bottom Yukawa leads to constructively interfering Higgs contributions to $B \to\tau\nu$ resulting in $R_{B\tau\nu} > 1$ as is actually preferred by the present data.
\end{itemize}

As discussed in section~\ref{sec:Btaunu}, for large $\tan\beta$ one expects strong constraints on the charged Higgs mass coming from $B\to\tau\nu$. In fig.~\ref{fig:Btaunu_MSSM} we show the BR$(B\to\tau\nu)$ as a function of the charged Higgs mass for $\tan\beta=(50,100,150,200)$. All the shown points fulfill the constraints from the other flavour observables as well as from $(g-2)_\mu$. The solid black curve corresponds to the result in absence of threshold corrections and the lower bound on the branching ratio in~(\ref{eq:RBtaunu_exp}) is shown as dashed line. We remind that the light Higgs mass region with $R_{B\tau\nu} > 0.51$ is actually excluded by data on the $K \to \mu \nu$ and $B \to D \tau \nu$ decays and at tree level the bound on the Higgs mass~(\ref{eq:Higgs_bound}) holds.
However, in particular for large $\mu$, sizable threshold corrections to $y_b$ lead to deviations from the tree level expectation. For positive values of $\mu$, the bound on the Higgs mass is relaxed. In our framework we find 
\begin{equation}
m_{H^\pm} \gtrsim (400,700,1200,1500) ~~\text{for}~ \tan\beta = (50,100,150,200)~.
\end{equation}

As $m_{H^\pm} \simeq m_{A^0}$, these still rather stringent bounds then do not allow for order of magnitude enhancements of BR$(B_s \to \mu^+\mu^-)$ anymore. Due to our limited range for $A_t$ in~(\ref{eq:LE_ranges}), we find that BR$(B_s \to \mu^+\mu^-)$ typically does not exceed $1 \times 10^{-8}$. Correspondingly, the so-called double Higgs penguin contributions to $B_s$ mixing~\cite{Buras:2002vd} do not lead to visible effects in the $B_s - \bar B_s$ mass difference $\Delta M_s$.

\section{General Gauge Mediation and uplifted SUSY}
\label{sec:ggm}

Gauge mediation of supersymmetry breaking\footnote{See \cite{Giudice:1998bp} for a review and collection of references.} is among the most successful ideas to solve the SUSY flavour problem. Since gauge interactions are flavour-blind, gauge mediation automatically leads to flavour-blind soft terms, and the possibility to have a mediation scale much lower than the Planck scale reduces also those flavour violating terms which are generated radiatively. While lots of different models of gauge mediation are on the market, a very simple and general definition of gauge mediation was suggested recently \cite{Meade:2008wd}: General Gauge Mediation (GGM) is a theory which decouples into the MSSM and a separate SUSY breaking hidden sector in the limit that the MSSM gauge couplings $g_i\to0$.

While this simple definition gives rise to well-known ingredients of gauge mediation, like small trilinear terms and flavour-blind soft terms at the mediation scale, it does not allow the generation of the $B\mu$ term in the soft SUSY breaking Lagrangian or the generation of a $\mu$ term from soft SUSY breaking interactions, since these parameters violate the Peccei-Quinn symmetry and cannot be generated by gauge interactions alone. To cure this problem, one possibility is to extend the definition of GGM to allow for interactions generating sufficient $\mu/B\mu$ \cite{Meade:2008wd}. Another possibility is to retain the original definition, implying $B\mu=0$ at the mediation scale, and assuming that the origin of the superpotential parameter $\mu$ is unrelated to the SUSY breaking sector.

This ``pure GGM'' setup with $B\mu=0$ at the mediation scale (cf. \cite{Abel:2009ve}) is not only interesting from a conceptual point of view, being a prediction of the simple definition of GGM, but also has an important phenomenological motivation: it strongly ameliorates the SUSY CP problem within gauge mediation.

The reason is that, in the MSSM with flavour-blind soft terms, CP-violating phases -- apart from the CKM phase and the strong CP phase -- can occur in the trilinear $A$ terms, the gaugino masses, the $B\mu$ and the $\mu$ term. Two of these phases can be rotated away by making use of the $U(1)_R$ and $U(1)_\text{PQ}$ symmetries, which are exact if the aforementioned parameters vanish. Since the $A$ terms are tiny at the gauge mediation scale, one can e.g. make one gaugino mass real as well as the $\mu$ term or the $B\mu$ term, but not both.
Assuming the gaugino masses to have a common phase at the mediation scale, as is the case for example in the Minimal Gauge Mediation (MGM) model, where the gaugino masses fulfill the GUT relation $M_1/g_1^2=M_2/g_2^2=M_3/g_3^2$, the condition $B\mu=0$ completely solves the CP problem, since there are no CP-violating phases left that cannot be rotated away.

\subsection{GGM parameter space and large \texorpdfstring{$\tan\beta$}{tan(beta)}}

It was shown in \cite{Meade:2008wd} that, given the definition of GGM, all the information on the hidden sector required to parametrize the soft sector is encoded in correlation functions of symmetry currents. Consequently, the soft terms at the mediation scale can be written to leading order as \cite{Buican:2008ws}
\begin{align}
M_k &= g_k^2 M B_k ~,& m_f^2 &= g_1^2 Y_f \zeta + \sum_k g_k^4 C_2(f,k) A_k ~,& A_{u,d,l}^{IJ} &= 0 ~,
\label{eq:GGM1}
\end{align}
where $C_2$ is the quadratic Casimir with respect to gauge group $k$ (we use the GUT normalization for $g_1$ and $C_2(f,1)=\frac{3}{5}Y_f^2$), $Y$ is the hypercharge and $m_f^2$ refers to the sfermion as well as Higgs squared masses.

We assume that relations (\ref{eq:GGM1}) are also valid for the Higgs soft masses $m_{H_{u,d}}^2$.
On the one hand, it is often assumed that these relations are less reliable for Higgses than for sfermions due to the possible impact of the physics responsible for curing the $\mu/B\mu$ problem. On the other, in the pure GGM setting where $B\mu=0$ at the mediation scale and $\mu$ is unrelated to the hidden sector, this assumption appears to be well justified.

In total, the GGM parameter space then consists of 4 real parameters $A_k$, $\zeta$ in the sfermion sector, 3 complex ones $B_k$ in the gaugino sector (in fact, only two of the three phases are physical as mentioned above), as well as $\tan\beta$ and the phase of $\mu$. If $B\mu=0$ at the mediation scale, $\tan\beta$ turns into a prediction and $\mu$ can be made real, with only the sign of $\mu$ still being a free parameter.

Since the seven sfermion and Higgs soft masses are determined by the four real parameters $A_{1,2,3}$ and $\zeta$, there are three relations among them, valid at the mediation scale $M$:
\begin{align}
6 m_Q^2+3m_U^2-9m_D^2-6m_L^2+m_E^2 &=0 ~,
\label{eq:GGM2-1}\\
m_L^2 &= m_{H_d}^2 ~,
\label{eq:GGM2-2}\\
3 m^2_Q - 3 m^2_U  + 3 m^2_L+2 m^2_E  &=6 m^2_{H_u} ~.
\label{eq:GGM2-3}
\end{align}

The hypercharge D-term $\zeta$ in (\ref{eq:GGM1}) gives rise to a non-positive definite contribution to the scalar squared masses and can lead to tachyons in the spectrum. To avoid this, it can be forbidden by imposing a discrete symmetry \cite{Dimopoulos:1996ig, Meade:2008wd}. However, since the two Higgs doublets differ in gauge quantum numbers only by the sign of their hypercharge, $\zeta=0$ would imply degenerate soft masses for the Higgs doublets, i.e. the three relations (\ref{eq:GGM2-1})--(\ref{eq:GGM2-3}) would be supplemented by a fourth one, $m_{H_u}^2 = m_{H_d}^2$.
As argued in section~\ref{sec:higgssoft}, such condition is problematic in the very large $\tan\beta$ regime since it obstructs radiative EWSB. Therefore, we will keep $\zeta$ nonzero in the following.

In fact, from a purely phenomenological perspective, the parameters $A_1$, $A_2$ and $\zeta$ in (\ref{eq:GGM1}) can be traded for $m_L^2\equiv m_{H_d}^2$, $m_{H_u}^2$ and $m_E^2$, which can be viewed as independent free parameters in that case. This parametrization is most transparent to avoid tachyonic sleptons from the outset. The squark masses are then determined in terms of these three parameters and $A_3$, and tachyonic squarks can in principle always be avoided by choosing $A_3$ sufficiently large.

It is important to note that, while a nonzero hypercharge D-term and thus a non-zero Higgs splitting allows to obtain a heavier Higgs spectrum by means of eq. (\ref{eq:DmH-1}), $m_{H_d}^2$ has to be positive at the mediation scale to avoid a charge breaking vacuum since it is tied to the left-handed slepton masses by means of (\ref{eq:GGM2-2}). Combining eqs. (\ref{eq:DmH-1}) and (\ref{eq:DmH-2}), one thus obtains an upper bound on the magnitude of $\mu$,
\begin{equation}
|\mu|^2 < m_A^2 + \frac{1}{2}m_Z^2 ~ ,
\label{eq:DmH-3}
\end{equation}
which is valid at the mediation scale, but should be expected to still hold approximately at low energies if the mediation scale is low.

\subsection{Numerical analysis of GGM at large \texorpdfstring{$\tan\beta$}{tan(beta)}}
\label{sec:ggmnum}

We now study the viability of the GGM setup in the very large $\tan\beta$ regime. Since we found in section~\ref{sec:lowenergy} that the perturbativity scale $\mu_\text{max}$ decreases fastly when $\tan\beta>50$, we will assume a relatively low mediation scale $M=100$~TeV throughout this section for definiteness.

We will assume the gaugino mass parameters $B_k$ in (\ref{eq:GGM1}) to be independent of the gauge group, implying the usual GUT-relations for gaugino masses. The reason for doing so is twofold; First, the solution to the CP problem in the uplifted SUSY scenario sketched above is facilitated if gaugino masses arise from a single scale as in the MGM. Second, as shown in section~\ref{sec:lowenergy}, $M_2<0$ requires $\mu<0$ to meet the $(g-2)_\mu$ bound, which in turn however leads to large positive threshold corrections to $y_b$ and lowers the scale until where $y_b$ remains perturbative. Therefore we restrict to the case with positive gaugino masses here.

We trade the parameters $A_1$, $A_2$ and $\zeta$ for $m_L^2$, $m_E^2$ and $\Delta m_H^2=m_{H_d}^2-m_{H_u}^2$ and scan in the following ranges,
\begin{align}
MB_k &\in [0, 2] ~\text{TeV} , &
A_3 &\in [0, 2] ~\text{TeV} , \\
m_L^2 &\in [0, 2] ~\text{TeV} , &
m_E^2 &\in [0, 2] ~\text{TeV} , &
\Delta m_H^2 &\in [0, 2] ~\text{TeV} . &
\end{align}

Note that $\Delta m_H^2\propto-\zeta$ and also that the above scan ranges imply negative values for $A_2$ in some cases, which however does not lead to tachyons if it is compensated by positive contributions from $A_1$ and $A_3$.

We first consider fixed values of $\tan\beta$, which requires treating $B\mu$ as a free parameter. A posteriori, we determine the according $B\mu$ terms at the mediation scale (as detailed in appendix~\ref{sec:Bmuapp}) to identify the regions in parameter space corresponding to the uplifted SUSY scenario with $B\mu(M)=0$.

\begin{figure}[tb]
\centering
\includegraphics[width=8cm]{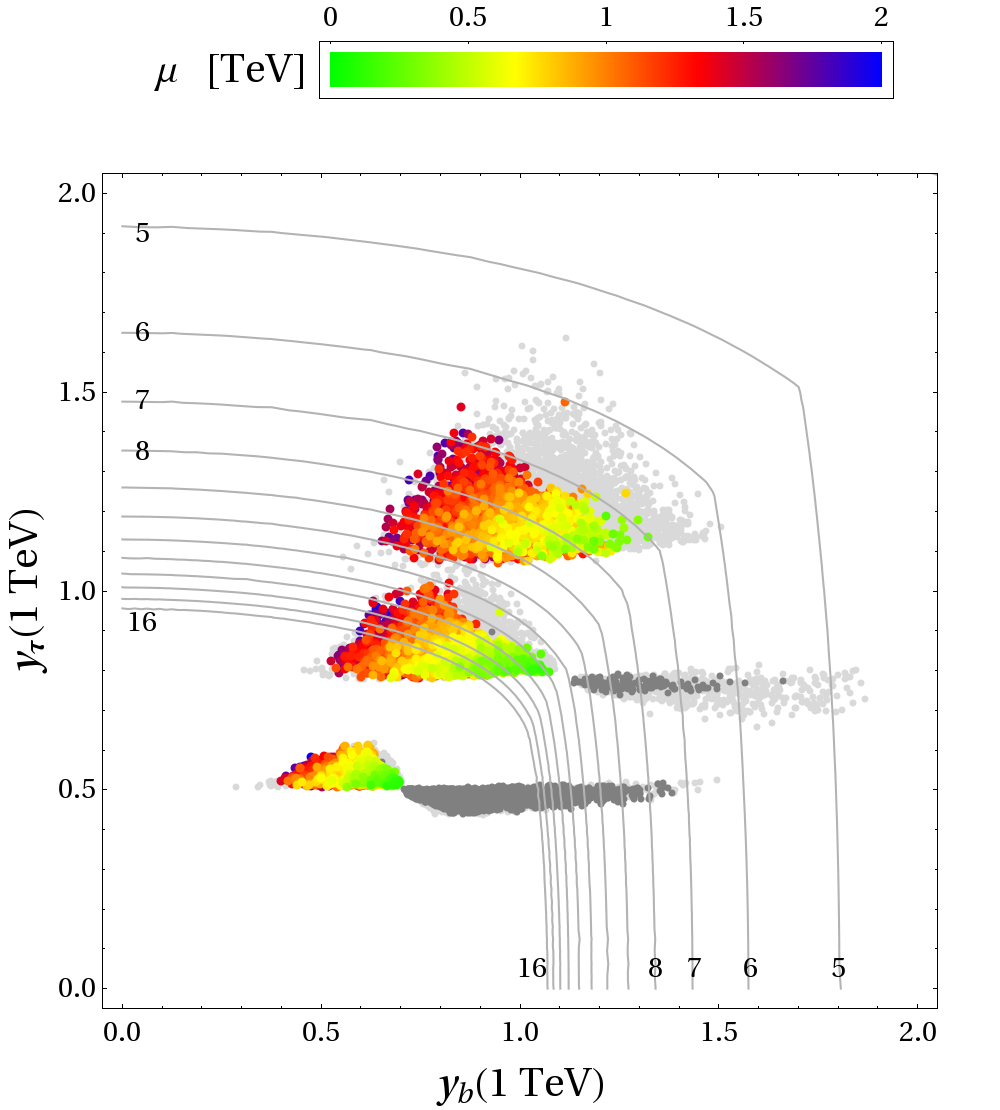}
\caption{Values of the bottom and tau Yukawas at the TeV scale in GGM with $\tan\beta=50,75,100$ (from bottom to top).
Light gray points are excluded at the $2\sigma$ level by the flavour constraints, while dark gray ones are excluded by $(g-2)_\mu$ alone at the $3\sigma$ level. Among the coloured points, the ones with smaller values of $\mu$ are stacked upon the ones with larger $\mu$.
The $\mu_\text{max}$ contours are as in fig.~\ref{fig:ybtaulimit}.
}
\label{fig:ybtau-GGM}
\end{figure}

Figure~\ref{fig:ybtau-GGM} shows the bottom and tau Yukawa couplings at the TeV scale for fixed $\tan\beta$ values 50, 75 and 100, akin to the plots for the low-energy scan in fig.~\ref{fig:ybtau-LE}. All the points with negative $\mu$ are ruled out at more than $3\sigma$ by the $(g-2)_\mu$ constraint. For $\tan\beta=100$, our code did not produce any converged points with correct EWSB for $\mu<0$. This can be understood from the proximity of the existing points to the $\mu_\text{max}=10^5$~GeV contour, signaling the non-perturbativity of one Yukawa coupling already at the mediation scale $M=10^5$~GeV, in which case the calculation becomes unreliable and the code unstable. An attempted scan with $\tan\beta=150$ accordingly produced no points at all, regardless of the sign of $\mu$. This is also in agreement with the $\tan\beta=150$ points with $\mu>0$ in the low-energy plot of fig.~\ref{fig:ybtau-LE}, where all points lie close to or beyond the $10^5$~GeV contour.

\begin{figure}[tbp]
\centering
\includegraphics[width=5.5cm]{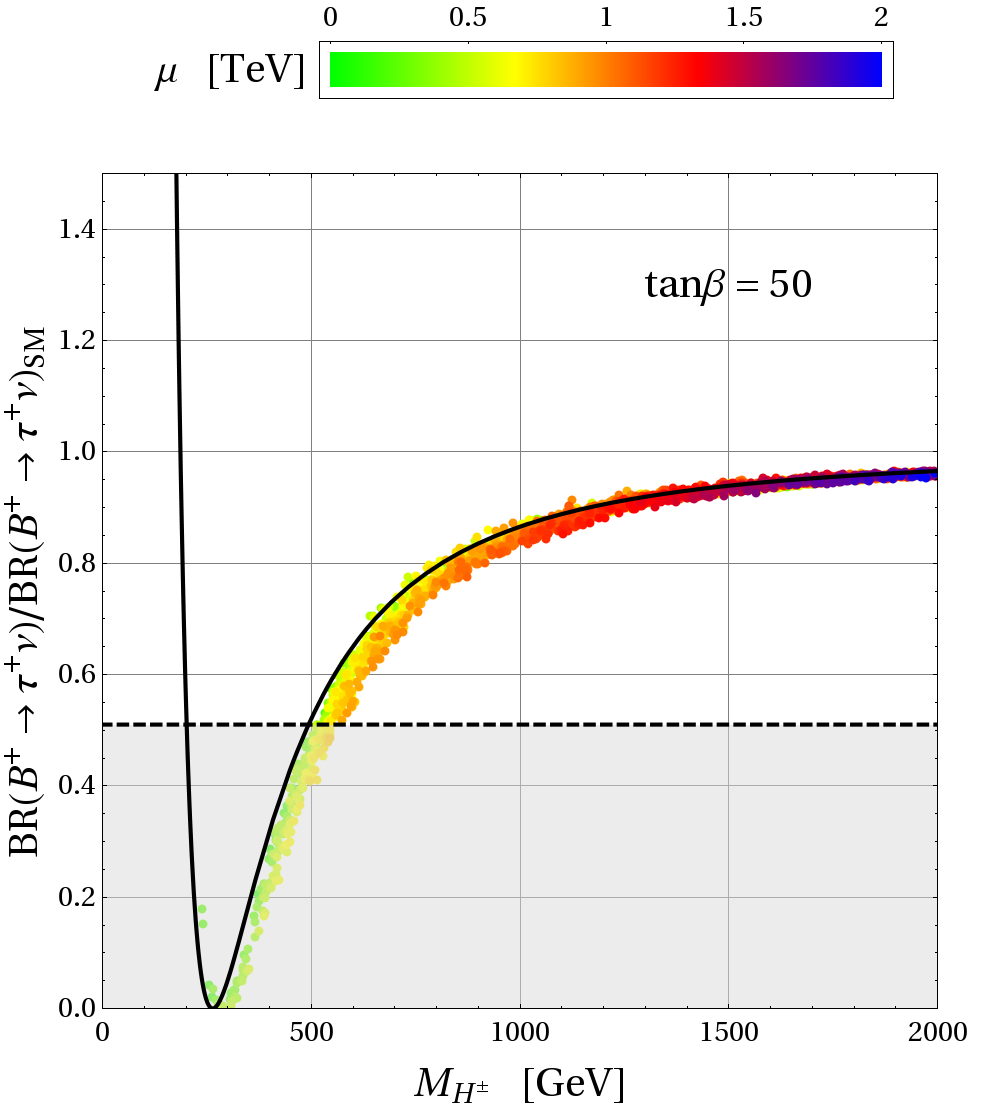}

\vspace{1em}
\includegraphics[width=5.5cm]{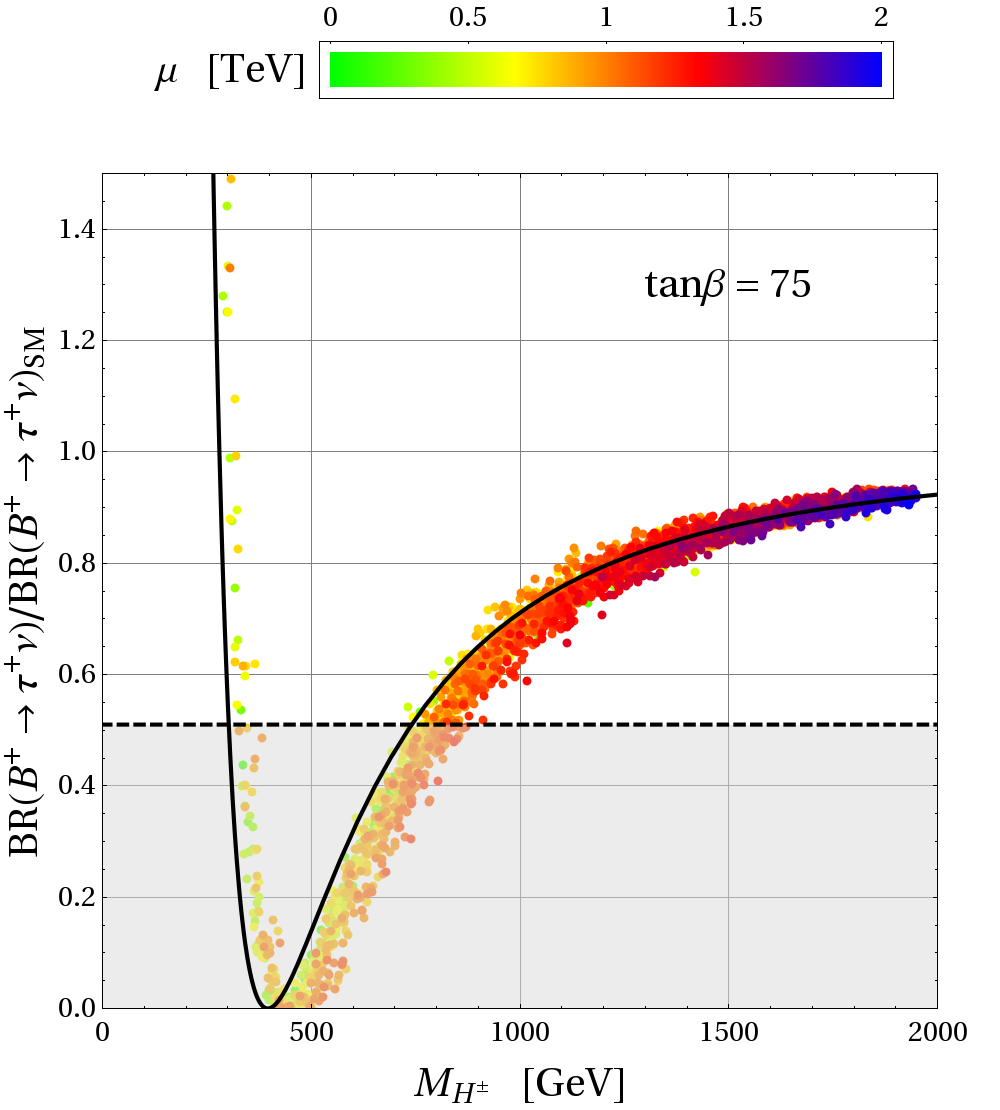}
\qquad
\includegraphics[width=5.5cm]{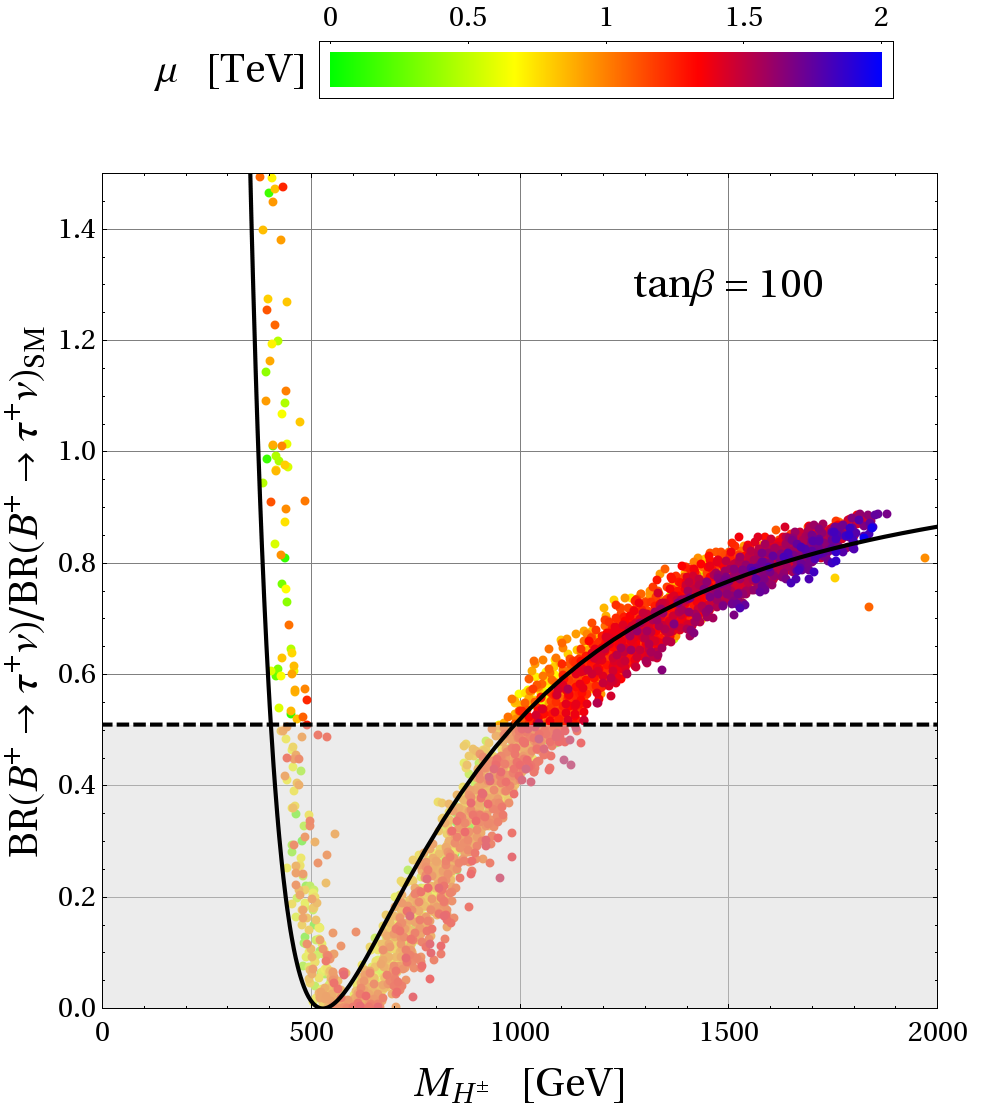}
\caption{Branching ratio of $B^+\to\tau^+\nu$ compared to its SM value as a function of the charged Higgs mass in GGM with $\tan\beta=50,75,100$.
All the shown points fulfill the constraints from the other flavour observables as well as from $(g-2)_\mu$. The solid black curve corresponds to the result in absence of threshold corrections and the experimental lower bound is shown as dashed line. Points with larger values of $\mu$ are stacked upon the ones with smaller $\mu$.
}
\label{fig:Btaunu_GGM}
\end{figure}

In fig.~\ref{fig:Btaunu_GGM} we show the BR$(B\to\tau\nu)$ as function of the charged Higgs mass for $\tan\beta = (50,75,100)$. We observe that the bound on the charged Higgs mass in the GGM scenario basically coincides with the one arising in absence of threshold corrections~(\ref{eq:Higgs_bound}). Approximately we find
\begin{equation}
m_{H^\pm} \gtrsim (500,750,1000) ~~\text{for}~ \tan\beta = (50,75,100)~.
\end{equation}
The reason for this at first sight surprising result is the upper bound on $|\mu|$ given in~(\ref{eq:DmH-3}). Whenever the Higgs mass is relatively low, also $\mu$ is small which excludes large positive threshold corrections to the bottom Yukawa $y_b$ that could relax the bound on the charged Higgs mass.

In fact, for small to medium values of $m_{H^\pm}$, negative corrections to the tau Yukawa are typically dominant, leading to slightly larger NP contributions to BR$(B\to\tau\nu)$ compared to the case without threshold corrections.
The other way round, we stress that in the regions of parameter space that are preferred from the point of view of Yukawa perturbativity, i.e. where both $y_b$ and $y_\tau$ are relatively small, the Higgs mass is always very heavy in GGM and the $B\to\tau\nu$ constraint is automatically fulfilled.
This can be understood by combining eqs.  (\ref{eq:DmH-1}), (\ref{eq:DmH-2}) and (\ref{eq:GGM2-2}) to obtain a relation between the $\mu$ term, the left-handed slepton mass and the pseudoscalar Higgs mass valid at tree level at the mediation scale,
\begin{equation}
m_A^2 = m_L^2 + |\mu|^2 - \frac{1}{2} m_Z^2 ~.
\label{eq:mAvsmL}
\end{equation}
Since the region with small $y_b$ and $y_\tau$ features heavy sleptons (to suppress the positive correction to $y_\tau$) and large $\mu$ (to enhance the negative corrections to $y_b$), eq.~(\ref{eq:mAvsmL}) implies that it also features a heavy Higgs spectrum.

Given the heavy Higgs spectrum and the fact that the trilinear coupling $A_t$ is naturally small in GGM as it is only generated radiatively in the running from the mediation scale $M = 10^5$~GeV down to the EW scale, we find only moderate effects in $B_s\to\mu^+\mu^-$, with BR$(B_s\to\mu^+\mu^-)$ typically not exceeding $5 \times 10^{-9}$. We even find many points that have values for BR$(B_s\to\mu^+\mu^-)$ below the SM prediction. In fact, charged Higgs contributions to $B_s\to\mu^+\mu^-$~\cite{Logan:2000iv} -- despite the fact that they are only proportional to $\tan^2\beta$ at the amplitude level -- become non-negligible in the considered scenario and can suppress the BR$(B_s\to\mu^+\mu^-)$ down to roughly $2\times 10^{-9}$.
We also find that $\Delta M_s$ remains essentially SM like.

In the introduction, we mentioned a vanishing $B\mu$ term at the mediation scale as a motivation for large or very large $\tan\beta$ (the uplifted SUSY scenario \cite{Dobrescu:2010mk}). If $B\mu=0$ at the mediation scale, $\tan\beta$ is no longer a free parameter but a prediction. Conversely, if $\tan\beta$ is fixed as in our GGM parameter scan, $B\mu$ at the mediation scale turns into a prediction and is not necessarily zero.
At tree level, one has $B\mu=m_A^2/\tan\beta$ at low energies.
At the loop level, this relation is corrected by the tadpole diagrams and the RG evolution as described in appendix~\ref{sec:Bmuapp}.
Still, the relation $B\mu(M)=0$ generically requires small pseudoscalar Higgs masses.

\begin{figure}[tb]
\centering
\includegraphics[width=1.95cm]{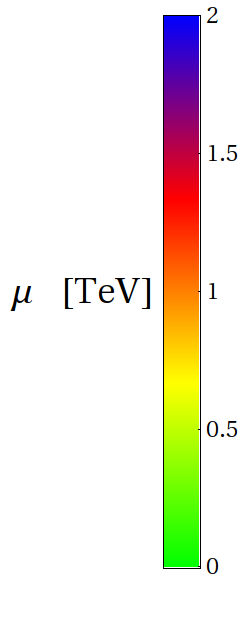}
\qquad
\includegraphics[width=5.5cm]{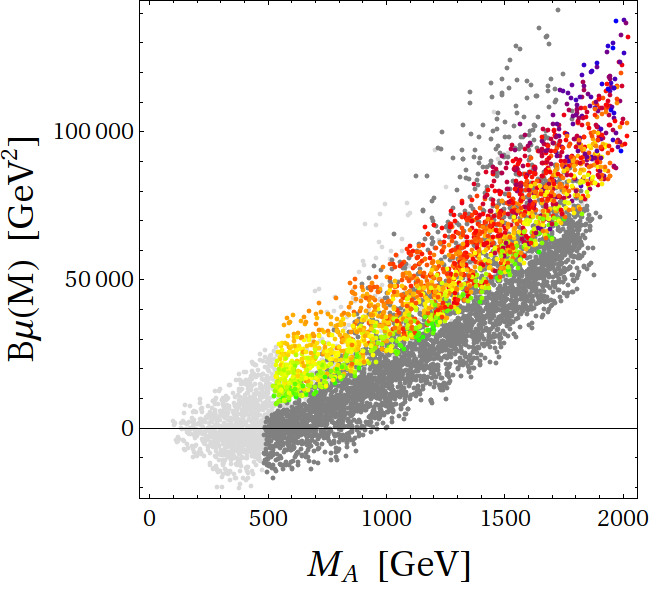}

\vspace{1em}
\includegraphics[width=5.5cm]{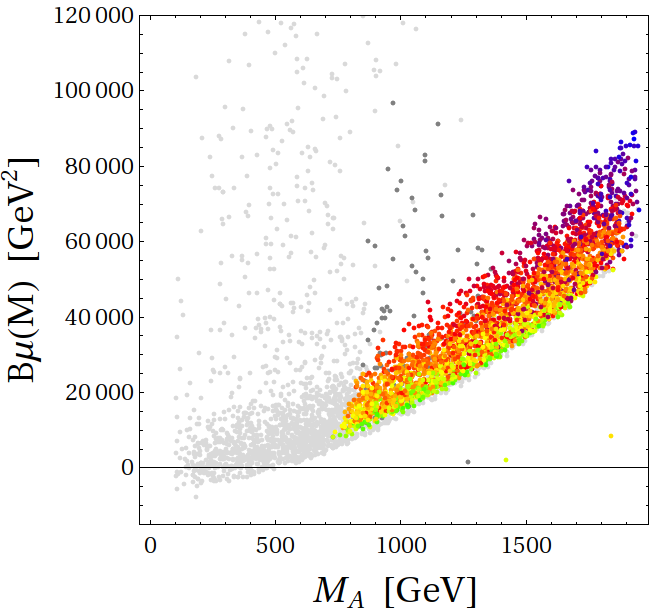}
\qquad
\includegraphics[width=5.5cm]{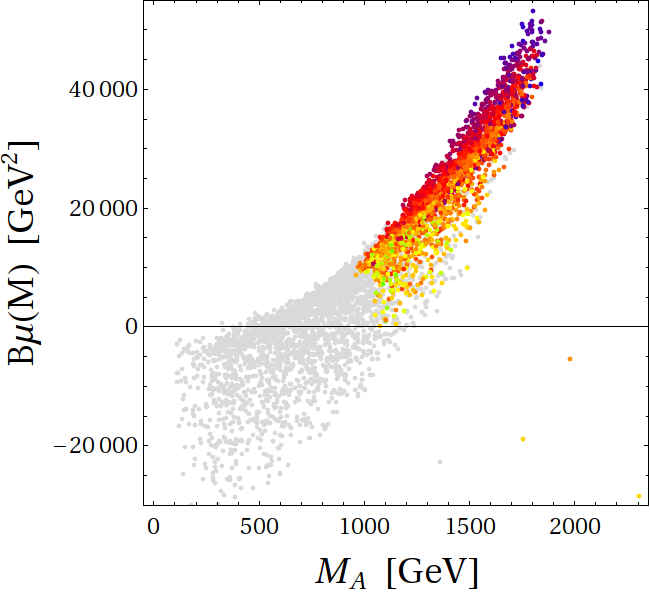}
\caption{$B\mu$ term at the mediation scale $M=100$~TeV vs. the pseudoscalar Higgs mass $m_A$ for $\tan\beta=50$ (top), 75 (bottom left) and 100 (bottom right) in the GGM scenario. The light gray points are ruled out by the flavour constraints at the $2\sigma$ level, while dark gray ones are excluded by $(g-2)_\mu$ alone at the $3\sigma$ level. In particular, the points for low $m_A$ are ruled out by $B\to\tau\nu$.
}
\label{fig:Bmu-vs-MA}
\end{figure}

In figure~\ref{fig:Bmu-vs-MA}, we show the $B\mu$ term at the mediation scale $M$ against the pseudoscalar Higgs mass $m_A$. As should be expected from the discussion in appendix~\ref{sec:Bmuapp}, the loop corrections to the tree-level relation $m_A^2=B\mu\,\tan\beta$ roughly scale with $\mu$. In particular, the lower half of the band at $\tan\beta=50$ corresponds to the points with negative $\mu$ and is ruled out at more than $3\sigma$ by the $(g-2)_\mu$ constraint. At low $m_A\lesssim500$~GeV, due to $m_A\approx m_{H^\pm}$, the $B\to\tau\nu$ constraint becomes active and rules out also the points with positive $\mu$. As a consequence, none of the points with $B\mu=0$ survives.

For $\tan\beta=75$, the points with $\mu<0$ have very large $y_b$ (cf. fig.~\ref{fig:ybtau-GGM}), leading to large loop corrections to $B\mu$. These scattered points are however also ruled out by $(g-2)_\mu$. At $m_A\lesssim750$~GeV, the points are ruled out by $B\to\tau\nu$. For $\tan\beta=100$, where all points have $\mu>0$, the bound on $m_A$ rises to almost 1~TeV.

In effect, the combined constraints from $(g-2)_\mu$ and $B\to\tau\nu$ make it virtually impossible to obtain viable scenarios in GGM with $B\mu=0$ at a mediation scale of 100~TeV.

With a higher mediation scale, $B\mu=0$ remains a valid possibility (cf. \cite{Abel:2009ve}) since the larger RG effects lead to stronger deviations from the tree-level relation $B\mu=m_A^2/\tan\beta$. However, a higher mediation scale also lowers the maximal allowed $\tan\beta$ value. This is clear from the discussion in section~\ref{sec:lowenergy}, which demonstrated that the maximum perturbativity scale $\mu_\text{max}$ quickly decreases for $\tan\beta\gg50$. Indeed, a trial scan within GGM with $M=10^7$~GeV and $\tan\beta=100$ did not return any points with perturbative Yukawa couplings.

We thus conclude that the region where $B\mu=0$ and $50<\tan\beta\lesssim100$ are simultaneously realized corresponds to a very restricted range in $M \approx 10^6 - 10^7$~GeV, below which the constraints from $B\to\tau\nu$ and/or $(g-2)_\mu$ are violated, and above which Yukawa couplings become non-perturbative. A full and systematic analysis of this parameter space is however beyond the scope of our present study.

\section{Conclusions} \label{sec:conclusions}

In this paper, we have studied constraints on the MSSM with $\tan\beta>50$ coming from the perturbativity of Yukawa couplings and from flavour physics. After reviewing the main constraints on the parameter space, we have performed a numerical analysis of the MSSM with flavour-blind soft terms at low energies and $50\leq\tan\beta\leq200$. We find that
\begin{itemize}
\item The $(g-2)_\mu$ constraint strongly disfavours positive contributions to the tau lepton mass from threshold corrections. From the point of view of perturbativity of Yukawa couplings, it is thus preferable that these contributions are small.
\item $B\to\tau\nu$ is a crucial constraint even for conservative estimates of the SM uncertainties. Interestingly enough, it disfavours the points with large $b$ and $\tau$ Yukawa couplings, leaving room for viable points with large $\tan\beta$ and not too low non-perturbativity scale.
\item The $B\to X_s\gamma$ and $B_s\to\mu^+\mu^-$ constraints can be relaxed if the trilinear couplings are small, as e.g. in models with gauge mediation.
\item The overall size of the threshold corrections to Yukawa couplings is limited by the size of $|\mu|$, which is in turn limited by the light Higgs mass bound and the requirement not to have a tachyonic stau or sbottom.
\end{itemize}

We have also analyzed the very large $\tan\beta$ region of General Gauge Mediation (GGM) with a low mediation scale. We have performed a numerical analysis of the parameter space of GGM with a mediation scale $M=100$~TeV for $\tan\beta=50$, 75 and 100, assuming GUT-like gaugino masses and a non-zero hypercharge D-term, that is required to obtain correct EWSB for very large $\tan\beta > 50$. In this setup we find that
\begin{itemize}
\item $\tan\beta\gtrsim150$ is not viable, since the Yukawas become non-perturbative below the mediation scale.
\item For $\mu<0$, all points are ruled out by $(g-2)_\mu$.
\item For $\mu>0$, the points with the highest non-perturbativity scale (i.e. smallest $y_\tau$) are automatically safe from the $B\to\tau\nu$ constraint since GGM relates the slepton masses and the Higgs soft masses such that for large slepton masses also the charged Higgs mass is large.
\item The lower bound on the charged Higgs mass from BR($B\to\tau\nu$) is much sharper than in the low-energy analysis due to the relation $\mu\lesssim m_A$ valid in the GGM setup.
\item $B\mu=0$ at the mediation scale is only possible for a negative $\mu$ parameter or low Higgs masses and correspondingly {strongly} disfavoured by the combined $(g-2)_\mu$ and BR($B\to\tau\nu$) constraints.
\end{itemize}
{
As stressed above,
the latter statement holds in the GGM setup described at the beginning of section~\ref{sec:ggmnum} with a very low mediation scale $M=100$~TeV and $\tan\beta\gtrsim50$.
With a higher mediation scale, $B\mu=0$ ist still possible,
but then the maximum allowed value for $\tan\beta$ quickly decreases.
}

\section*{Acknowledgments}

We thank Andrzej Buras, Bogdan Dobrescu, Patrick Fox and Paride Paradisi for useful discussions and comments on the manuscript.
This work was supported by the Cluster of Excellence ``Origin and Structure of the Universe'', the German Bundesministerium f\"ur Bildung und Forschung under contract 05HT6WOA and the Graduiertenkolleg GRK 1054 of DFG. We also thank the Galileo Galilei Institute for Theoretical Physics for the hospitality and the INFN for partial support during the completion of this work.

\appendix

\section{Determining the \texorpdfstring{$B\mu$}{Bmu} term at the mediation scale}
\label{sec:Bmuapp}

In this appendix, we describe how to obtain the value of $B\mu$ at some arbitrarily high energy scale, starting from an MSSM parameter point at low energies\footnote{A similar calculation has also been presented in \cite{Dobrescu:2010mk}.}. The same procedure can of course be applied vice versa to calculate the low-scale $B\mu$ term (and predict $\tan\beta$) starting from some high-scale value (e.g. $B\mu(M)=0$). In the latter case however, one has to perform an iteration since the loop corrections to $B\mu$ depend on the value of $\tan\beta$.

The $B\mu$ term at low energies is determined from the EWSB conditions,
\begin{align}
 \frac{m_Z^2}{2} \cos(2\beta) + m_{H_d}^2 + |\mu|^2 - B\mu \tan\beta &= -\frac{T_d}{v_d} ~,
\label{eq:ewsb1}\\
-\frac{m_Z^2}{2} \cos(2\beta) + m_{H_u}^2 + |\mu|^2 - B\mu \cot\beta&=  -\frac{T_u}{v_u} ~,
\label{eq:ewsb2}
\end{align}
where $T_{u,d}$ are the tadpole corrections arising at one-loop level. These conditions are usually imposed at the scale $m_\text{SUSY}=\sqrt{m_{\tilde t_1}m_{\tilde t_2}}$  to minimize logarithmic contributions to $T_{u,d}$ from stop loops. However, bottom, stau, chargino or neutralino contributions can still be sizable.

From (\ref{eq:ewsb1}) and (\ref{eq:ewsb2}) one finds
\begin{align}
B\mu \tan\beta &= -m_Z^2 +  m_{H_d}^2 - m_{H_u}^2 - \frac{T_d}{v_d} + \frac{T_u}{v_u}  + O(1/\tan^2\beta)
\nonumber\\&
\approx m_A^2 - \Delta T \tan\beta ~,
\end{align}
where we have defined $T_d=\Delta T \,v_u + O(v_d)$. In the Mass Insertion Approximation (MIA), one finds
\begin{align}
16 \pi^2 ~ \Delta T =\ &
+ 3 \mu y_t^2 A_t  ~f_3(x_{\tilde t_L},x_{\tilde t_R})
+ 3 \mu y_b^2 A_b  ~f_3(x_{\tilde b_L},x_{\tilde b_R})
+ \mu y_\tau^2 A_\tau ~f_3(x_{\tilde \tau_L},x_{\tilde \tau_R})
\nonumber\\&
+ 3 g_2^2 \mu M_2  ~f_3(x_\mu,x_2) + \frac{3}{5} g_1^2 \mu M_1  ~f_3(x_\mu,x_1) ~,
\label{eq:tadppr}
\end{align}
where $x_\mu=\mu^2/m_\text{SUSY}^2$, $x_{1,2}=M_{1,2}^2/m_\text{SUSY}^2$ and $x_{\tilde f_{L,R}}=m_{\tilde f_{L,R}}^2/m_\text{SUSY}^2$. For the loop function, one has $f_3(1,1)=2$ and the precise form is given in appendix~\ref{sec:loop}.

The so obtained $B\mu$ term at the scale $m_\text{SUSY}$ can then be evolved to the mediation scale $M$ by means of its RG equation. At one loop, the RGE for $B=B\mu/\mu$ can be written as
\begin{equation}
\frac{dB}{dt} = \frac{1}{8\pi^2}
\left(  3 y_t^2 A_t  + 3 y_b^2 A_b+ y_\tau^2 A_\tau +3g_2^2\, M_2 +\frac{3}{5} g_1^2\, M_1\right)~.
\label{RGE:B}
\end{equation}
Note that the scale dependence of the tadpole (\ref{eq:tadppr}) is canceled in the leading log approximation to the one-loop RG running (\ref{RGE:B}).

In our numerical analysis, we take into account the full one-loop calculation of the tadpole contributions (without making use of the MIA) \cite{Pierce:1996zz} and use two-loop RG running for $B\mu$ (as for all the other parameters) \cite{Martin:1993zk}.

\section{Loop functions}\label{sec:loop}

\begin{eqnarray}
f_1(x) &=& \frac{1}{1-x} + \frac{x}{(1-x)^2} \log x ~, \\[12pt]
f_2(x,y)&=& \frac{x \log x}{(1-x)(y-x)} + \frac{y \log y}{(1-y)(x-y)}~, \\[12pt]
f_3(x,y) &=& 1 + \frac{x \log x}{x-y} - \frac{y \log y}{x-y}~,
\end{eqnarray}

\begin{eqnarray}
f_4(x,y) &=& \frac{8-25y+11y^2-3x^2(1+y)+x(1+16y-5y^2)}{2(1-x)^2(1-y)^2(x-y)} \nonumber \\
&& + \frac{7x^2-4x^3-4y+xy}{(1-x)^3(y-x)^2} \log x + \frac{(4-3x-5y+4y^2)y}{(1-y)^3(x-y)^2} \log y ~,
\end{eqnarray}

\begin{eqnarray}
f_7(x) &=& -\frac{13-7x}{12(1-x)^3} - \frac{3+2x-2x^2}{6(1-x)^4} \log x ~, \\[12pt]
f_8(x)&=& \frac{1+5x}{4(1-x)^3} + \frac{x(2+x)}{2(1-x)^4} \log x~, \\[12pt]
\tilde f_7(x) &=& \frac{4(1+5x)}{18(1-x)^3} + \frac{4x(2+x)}{9(1-x)^4} \log x ~, \\[12pt]
\tilde f_8(x)&=& -\frac{11+x}{3(1-x)^3} - \frac{9+16x-x^2}{6(1-x)^4} \log x~.
\end{eqnarray}

\bibliography{very_large_tanb}
\bibliographystyle{utphys}

\end{document}